\documentclass[reprint,superscriptaddress,amsmath,amssymb,aps,pra]{revtex4-2}

\usepackage{xcolor,graphicx}
\usepackage{amssymb,amsmath,bbm}
\usepackage[utf8]{inputenc}
\usepackage{comment}
\usepackage{braket}
\usepackage[normalem]{ulem}
\usepackage{dcolumn}
\usepackage{bm}

\newcolumntype{K}[1]{>{\centering\arraybackslash}p{#1}}

\begin{document}
	\title{Coexistence of local and nonlocal shock waves in nanomaterials}
	
	\author{Ilaria Gianani}
	\thanks{These authors contributed equally}
	\affiliation{Dipartimento di Scienze, Università degli Studi Roma Tre, Via della Vasca Navale 84, 00146, Rome, Italy}
	
	\author{Silvia Gentilini}
	\thanks{These authors contributed equally}
	\affiliation{Istituto dei Sistemi Complessi -- CNR, Piazzale Aldo Moro 5, 00185, Rome, Italy}
	
	\author{Iole Venditti}
	\affiliation{Dipartimento di Scienze, Università degli Studi Roma Tre, Via della Vasca Navale 84, 00146, Rome, Italy}
	\author{Chiara Battocchio}
	\affiliation{Dipartimento di Scienze, Università degli Studi Roma Tre, Via della Vasca Navale 84, 00146, Rome, Italy}
	\author{Neda Ghofraniha}
	\email{These authors contributed equally}
	\affiliation{Istituto dei Sistemi Complessi -- CNR, Piazzale Aldo Moro 5, 00185, Rome, Italy}
	\author{Marco Barbieri}
	\email{These authors contributed equally}
	\affiliation{Dipartimento di Scienze, Università degli Studi Roma Tre, Via della Vasca Navale 84, 00146, Rome, Italy}
	\affiliation{Istituto Nazionale di Ottica -- CNR, Largo Enrico Fermi 6, 50125 Florence, Italy}

	\begin{abstract}
     {\bf 
    \noindent High-performing nanomaterials are paramount for advanced photonic technologies, like sensing, lasing, imaging, data storage, processing, and medical and biological applications. Metal nanoparticles play a key role, because the localized surface plasmonic resonances enhance the linear and nonlinear optical properties of hosting materials, thus increasing the range of applications. The improved optical response results in an enriched and largely unexplored phenomenology. In the present work, we show the formation and interaction of two dispersive shock waves emerging by illuminating an aqueous suspension of gold nanorods with a pulsed laser beam. By analyzing the characteristic undular bores that regularize a shock front in a dispersive material, we were able to distinguish dispersive shock waves of different origins, i.e., electronic and thermal, first coexisting and then interacting to form a double front of equal intensity. The observed scenario agrees with existing literature reporting an electronic and thermal contribution to the nonlinear refractive index of material containing gold nanoparticles. Although a full comprehension of the reported results deserves an analytical description, they already pave the way to using materials containing metal nanoparticles as a platform for studying fundamental extreme nonlinear phenomena and developing novel solutions for sensing and control.}
    \end{abstract}
	\maketitle

	Engineered nanomaterials have become pivotal for developing innovative solutions in a large number of fields ranging from biomedical to environmental applications due to unique tailoring capabilities of their chemical and physical properties~\cite{Haine,AnkriFixler,Saha,Chen}. In particular, metal nanoparticles such as as gold nanorods (AuNRs), have emerged as precious tools for advanced biomedical applications, such as imaging, photothermal therapy, drug delivery, and biosensors ~\cite{app9163232,Deinavizadeh,Huang22}. When interacting with light, their size dictates that the response of AuNRs is governed by the confinement of the electric field, originating a localized surface plasmon resonance (LSPR). In the case of AuNRs, this exhibits two wide distinct bands in the visible-near infrared, where biological tissues are not active: this makes them excellent tools for diagnosis and therapy. 
	The presence of the LSPR is of paramount importance as it enables exploring resonant phenomena at ultrafast timescales, while enhancing the linear and nonlinear optical properties of the material~\cite{Sando07,Mayer11,Olesiak12,Park13,Furube17}.

	The landscape of the nonlinear optical response of metallic nanoparticles is further complicated by the influence of the environment in which the particles lie. This results in an interplay between two contributions: an electronic, local, nonlinear contribution dependent on the LSPR, and a thermal, nonlocal one depending on the environment~\cite{Zha17}. The double origin of the nonlinear response has been inferred by Z-scan measurements, mainly addressing self-focusing effects~\cite{Sou08,Kar10}. The phenomenology, however, is far richer and has remained hitherto unexplored, in particular for what concerns extreme events such as optical dispersive shock waves. 
	
	Here we show that when an aqueous solution of AuNRs is illuminated with a collimated pulsed laser at different intensities, the interaction between local and nonlocal Kerr-like defocusing nonlinearities leads to the formation and collision of local and nonlocal shock waves. We demonstrate by direct visualization and for the first time the coexistence and interaction of these two effects. Our results show colliding shock waves of different nature originating by the same laser profile, where the collision is not provoked by any beam-shaping tool, but it is rather auto-generated by the mutual interaction of the two distinct optical nonlinearities. Our investigations disclose yet another fundamental behaviour in the nonlinear response of AuNRs, ready to be harnessed for unprecedented applications, and providing additional control to current ones.
	
	\section*{Results}
	\noindent {\bf Dispersive shock waves.}
    An optical pulse propagating in the presence of a strong defocusing nonlinearity steepens and eventually results in a gradient catastrophe and wave breaking~\cite{Con09,Fat14}. This behaviour is often described by a hydrodynamical treatment in the limit of a dissipation-free, dispersive medium. This regime is characterised by the development of an oscillatory front, which is regularised by dispersion: the analogy is preserved in that diffraction acts as dispersion for light in stabilizing its propagation so that dispersive shock waves (DSW) are formed.
 
    The formal treatment introduced by Gurevich and Pitaeski relies on a Nonlinear Shr\"{o}dinger equation (NLSE). This has succeeded in describing disparate systems undergoing the same evolution, from shallow water and ion-acoustic waves~\cite{Per66,Tay70,Tri16,Low13}, to Bose-Einstein condensates~\cite{Mep09, Hoe06,Dut01,kam04,Dam04} and to light propagation in time through fibers \cite{Rot89, Kiv90,Yur93} and in space in nonlinear bulk media~\cite{Wan07,Gho07,Gen13,Xu15}. 
    
    For a slowing varying amplitude of an optical field $\psi$, the NLSE is written as
    \begin{equation}
    i\varepsilon\frac{\partial\psi}{\partial z}+\frac{\varepsilon^2}{2}\nabla^2_{x,y}\psi+\chi \theta\psi=0.
    \label{eq.NLS}\end{equation}
    This expression is obtained introducing rescaled quantities in the paraxial wave equation obeyed by the envelope $A$ of a monochromatic field.  In particular, $\Delta n=n_2 |A|^2$ is the nonlinear variation of the refractive index, and $n_2$ is the local Kerr coefficient. We have also defined
    $\chi=n_2/|n2|=\pm1$ as the sign of the nonlinearity, $\varepsilon^2=L_{\rm nl}/L_{\rm d}$ as the ratio of the nonlinear length $L_{\rm nl}$ to the diffraction length $L_{\rm d}$ and $\theta=k_0L_{nl}|\Delta n|$. The derivation is sketched in the Methods. 
    
    To reduce complexity and illustrate the basic physics, we can consider the one-dimensional case, thus $r\rightarrow x$. In the strongly non linear regime, i.e. $\varepsilon\ll 1$, the one dimensional version of Eq. (\ref{eq.NLS}) admits solutions in the form $\psi=\sqrt{\rho(x,z)}\exp{[i\phi(x,z)/\varepsilon]}$ and can be reduced to Euler-like fluid equations (see Methods): 
    \begin{equation}
\rho_z+(\rho u)_x=0,\\\label{eq.hydrodynamic1}
\end{equation}
\begin{equation}
u_z+uu_z-\chi\theta_x=0.
\label{eq.hydrodynamic2}
\end{equation}
where $u=\phi_x=d\phi/dx$ is the radial phase chirp.  In the defocussing case ($\chi=-1$) for an ideal medium ($\theta=\rho$), Eqs. (\ref{eq.hydrodynamic1}) and (\ref{eq.hydrodynamic2}) correspond to the same system of conservation laws that govern gas dynamics, with $u$ and $\rho$ being the velocity and mass density, respectively, of a gas at a pressure $\rho^2$~\cite{Gho07}.


     The shock manifests as the formation of undular bores, contrasting the generation of multivalued regions in the rapidly oscillating solutions. The position of these bores with respect to the shock front depends on the physical origin of $\Delta n$. For local nonlinearities, typical of electronic response, the formation of undular bores is external to the shock front, as shown numerically~\cite{Iso19} and experimentally by direct visualization at the output face of a photorefractive crystal with defocusing nonlinearity generated by a laser beam~\cite{Wan07}. Recently, the description of light oscillatory shocks by the Whitham modulation theory, relevant in the weakly dispersive regime~\cite{Iso19,Iva20}, has been validated in an experiment in a nonlinear gaseous medium~\cite{Bie21}.

    The behaviour of the oscillations changes in presence of nonlocal nonlinearity. Nonlocality plays a key role in many physical systems and many optical materials respond nonlocally, in that the variation of the index of refraction occurs also in a region surrounding the impinging laser beam and not only the exact point of excitation. In the context of nonlinear optics a variety of nonlocal mechanisms has been  addressed~\cite{Sut93,Yav05,Con03}. Barsi et al. showed that nonlocality in thermal nonlinear systems creates an effective damping force, inhibiting the shock speed and reducing oscillations~\cite{Bar07}. Other works have given experimental and numerical evidence of undular bores internal to the steep shock edge in  post-shock patterns in materials with thermal nonlocal nonlinearity~\cite{Gho07,Gho12,Mar19}.
	
	This classification helps as long as one of the two mechanism (local or nonlocal) prevails. When AuNRs interact with ultrashort pulsed lasers, however, it is not trivial to establish {\it a priori} which of the two nonlinearities dominates the dynamics: on the one hand, using spectrally broad pulses means ensuring a resonating behaviour, thus stimulating local nonlinearities due to the electronic response; on the other, nanorods are engineered to be optimal energy transductors converting electromagnetic energy to thermal, and as such it should be expected that using temporally short, highly energetic pulses should incur in a thermal response. The interplay between this twofold nature of AuNRs nonlinearities is known and has been studied extensively through Z-scan techniques~\cite{Sou08}, which have demonstrate a large nonlinearity enhancement of several orders of magnitude~\cite{Nan22}, making them ideal candidates for the direct visualization of both local and nonlocal shock waves. 
	
	In this work, we experimentally investigate the spatial propagation of a Gaussian laser mode subject to the two Kerr-like defocusing AuNRs nonlinearities just discussed: the local one of electronic origin and the nonlocal of thermo-optical origin.
	Remarkably, the enhancement in the nonlinear response of AuNRs allows to directly observe the shock fronts with a collimated beam, thus suggesting this system as a valuable platform for further investigations on extreme nonlinear phenomena. We expect to unearth the double origin of AuNRs shock waves employing two different approaches: by analyzing the temporal dynamics, we expect to observe two different timescales for the formation of local and nonlocal fronts; by exploring different power regimes, we expect to see a transition between a purely local effect at low powers, leading to the coexistence of the two phenomena at higher powers.\\
	
\begin{figure*}[]
	\includegraphics[width=\textwidth]{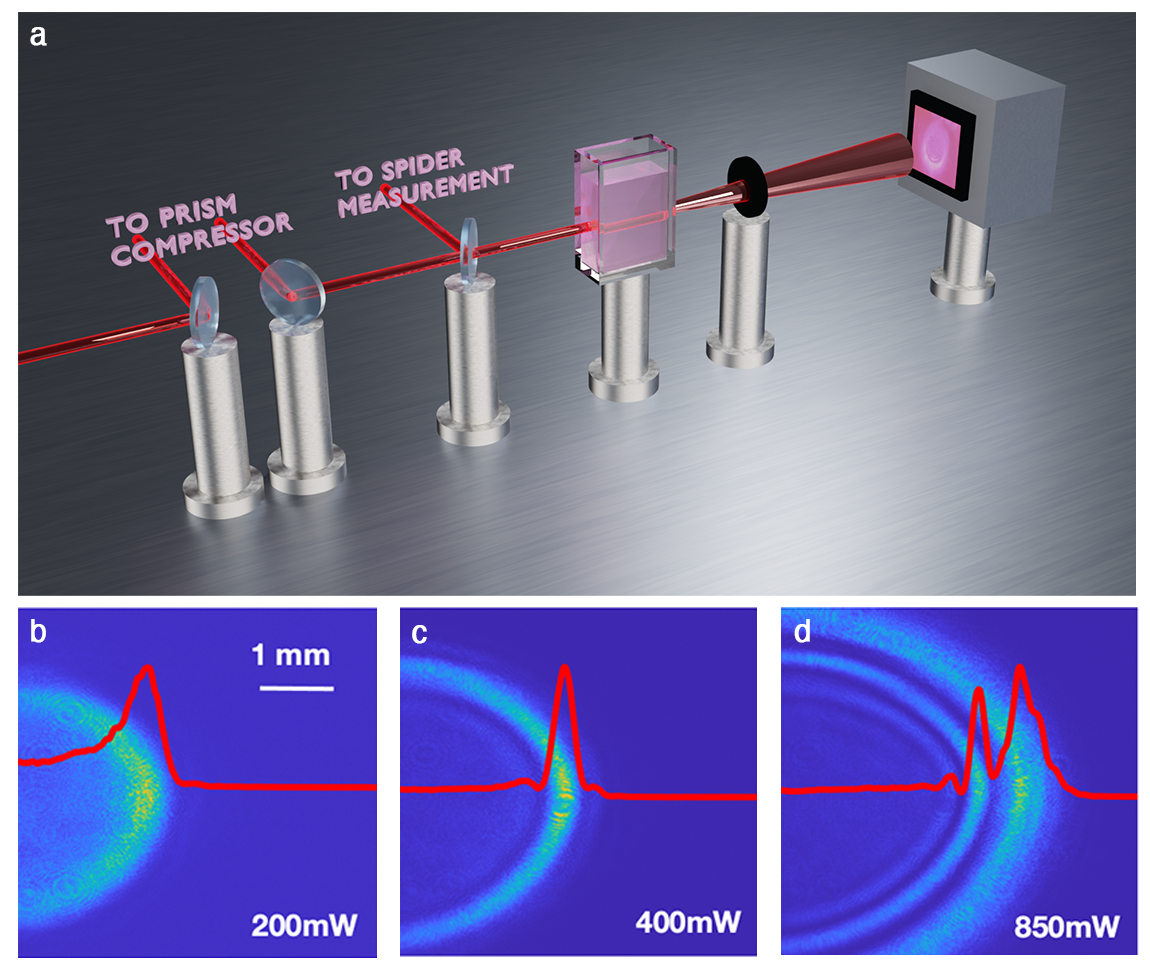}
	\caption [width=\textwidth]{(a) Sketch of the experimental setup. A 159 fs Ti:Saph laser centered at 807 nm interacts with a solution of AuNRs (pink cuvette). The transmitted beam is attenuated using absorptive ND filters and the spatial profile is captured using a CCD camera.  See the Methods for further details. (b-d) Intensity profiles of the transmitted beam as collected by the CCD camera at three different incident power $P=200,400,850$ mW. The superimposed curves are obtained by averaging the profiles from the images over 10 pixel rows.}
	\label{fig1}
	\end{figure*}

	{\bf \noindent Experiment.} We synthesise and prepare a solution of AuNRs (see Methods) which we place in a 0.8 cm wide quartz cuvette with a 2 cm path length along the beam direction. A sketch of the optical setup is illustrated in Fig. 1(a): a tunable Ti:Sapphire laser delivering 159 fs pulses centered at 807 nm illuminates the sample, and the transmission at the exit of the cuvette is imaged on a CCD camera without any focalizing optics. The experimental details are reported in the Methods section. The optical nonlinearity is regulated by the power of the input laser: at intensities below 50 mW the nonlinear response is negligible and the beam at the exit of the sample retains its Gaussian profile. By increasing the power a self-defocusing effect is observed. The transmitted beam first broadens abruptly and then in breaks into concentric undulations signature of optical dispersive shock waves.  In Figure~\ref{fig1} (b-d) we show as an illustrative example the recorded images with DSWs at three different input power $P=200$, 400, 850 mW. The superimposed red curves are the averages calculated over a slice with a ten pixels thickness. The problem is expected to be fully symmetric around the propagation direction. However a concentration gradient in the AuNRs solution occurs, resulting in a lower intensity in the upper half of the beam. We thus perform our analysis on the other half of the spatial mode. In all the analysis performed the position x indicates the distance from the center of the spatial mode. \\

	{\bf \noindent Shocks dynamics in time.} We perform the temporal measurements using the CCD camera, which allows us to track the dynamics of the spatial profiles  with a resolution of 30 fps. The measurements are performed as follows: we first set the power at the chosen value, and block the laser beam with a misaligned iris acting as a beam-block. We then start the acquisition and then subsequently open the iris. This allows us to fully capture the early dynamics within the limitations imposed by the temporal resolution. 
	
\begin{figure*}[h!]
    \includegraphics[width=\textwidth]{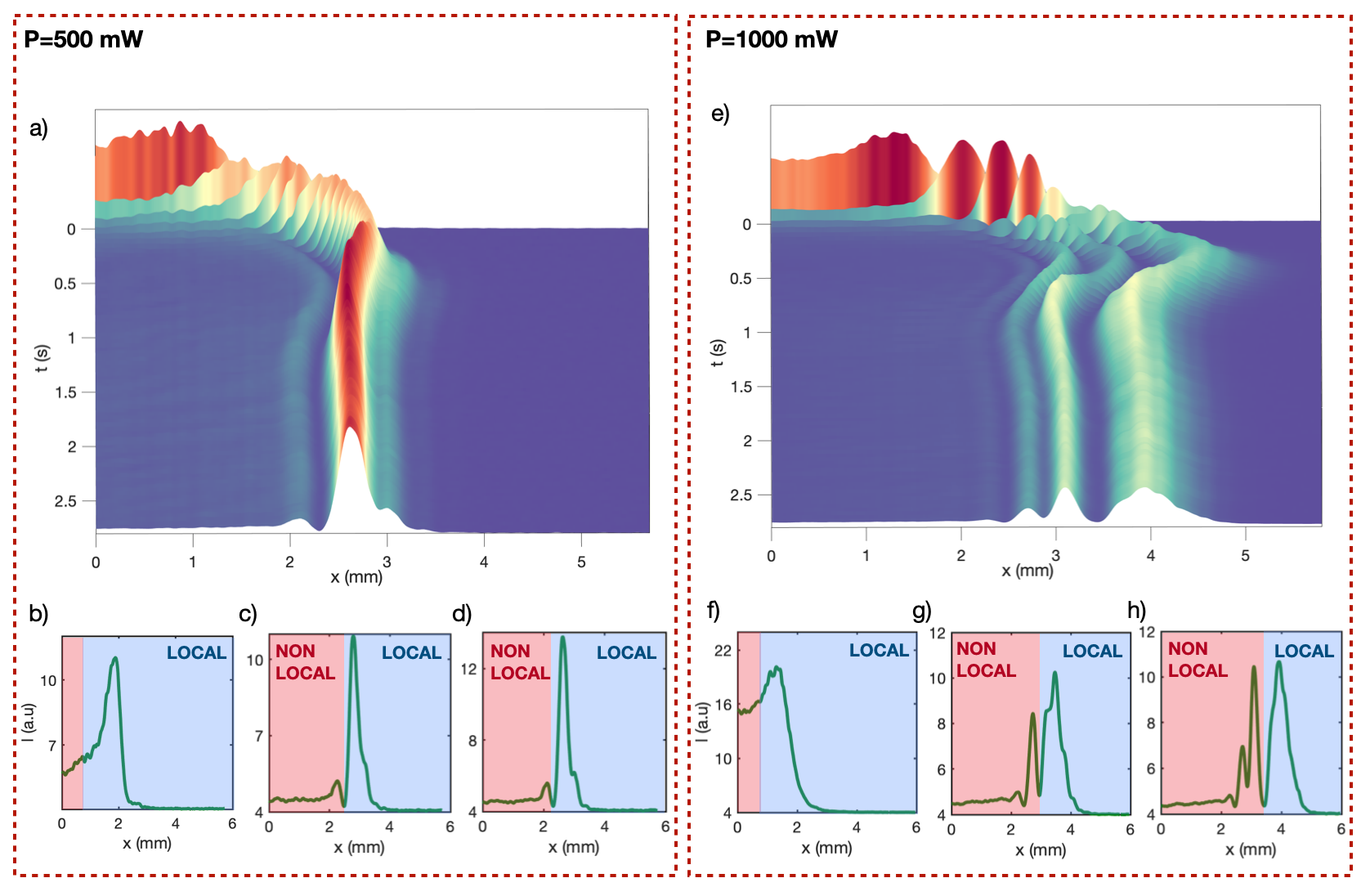}
    \caption {Time-space ($t$-$x$) map of the intensity profiles collected as a function of time at fixed power $P=0.5$ W (a) and $P=1$ W (e). The profiles are collected from time $t=0$ s to $2.8$ s with a time step equal to $1/30$s as described in the main text.  The bottom panels  display the spatial profiles for the two powers at times $t=0.03$ s (b); $0.6$ s (c); $2.1$ s (d); $0.09$ s (f); $0.18$ s (g), $1.89$ s (h). These show how at lower powers the dynamic of the local and nonlocal phenomena is characterized by their coexistence with the only influence being an incipit of the thermal lensing effect which contrasts the profile expansion. At higher powers the two shock waves fully interact with the nonlocal fronts slowly prevailing.}
    \label{fig:dynamics}
    \end{figure*}

	The results are shown in the maps of Fig. \ref{fig:dynamics} for $P=$0.5 W (a) and $P=$1.0 W (b) respectively, where the half-beam intensity of the spatial profile averaged over 10 pixel rows is reported as a function of time. At both powers we observe a transient phase of similar length $t=0.75$ s, after which the profile reaches a stationary state. At $P=0.5$ W, a predominant peak which emerges on time scales shorter than our resolution can capture, its intensity lowers during the transient phase, and then increases again and stabilizes during the stationary state. Given the early emergence of this front, we ascribe it to a local nonlinear response. At $t=0.3$ s, an internal peak emerges, signalling the presence of a nonlocal response. This second peak however never becomes dominant, however it disrupts the position of the electronic peak which undergoes a contraction. By $t=0,75$s, the profile reaches a stationary state in which the local peaks and  the thermal one coexist. For $P=1.0$ W the dynamics is more involved: it begins as in the previous instance with only the electronic fronts, then at about $t=0.09$ s an internal (thermal) oscillation appears, but does not yet prevail. We note how, due to the increased power, the thermal nonlocal response now happens earlier in the dynamics. At $t=0.15$ s the main and secondary electronic fronts drop and the main thermal one reaches an intensity comparable to the third electronic front. Also at this power we assist to a contraction of the fronts. At $t=0.75$ s, the intensities and positions of the fronts stabilize reaching a stationary state. We ascribe the behaviour at both powers to a collision between the electronic and thermal fronts: at $P=0.5$ W the disruption caused by the thermal front is minimal and only affects the position of the electronic fronts, thus denoting the onset of the collision between the electronic and the emerging thermal fronts; at $P=1$ W the collision is fully unfolding, and this results in the destruction of the main and secondary electronic front and the creation of a double front comprising the main thermal and third electronic fronts. In the double front regime, the internal nonlocal undular bores are well separated in space from the local external ones and, more importantly, they travel in parallel with comparable intensity.
	
\begin{figure*}[h]
    \includegraphics[width=\textwidth]{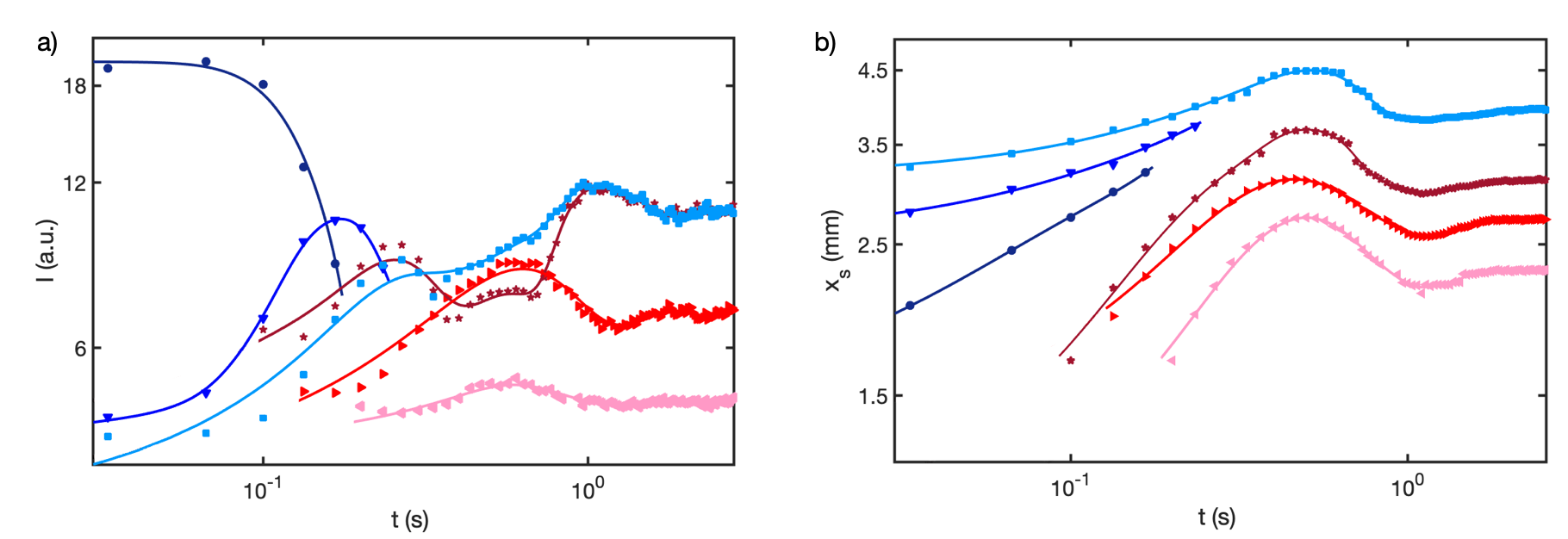}
    \caption {Characterization of the DSWs dynamics. (a) Log-log plot of the intensity of the electronic and thermal fronts as a function of time (b) Log-log plot of the position of the peaks of the electronic and thermal fronts as a function of time. The color scheme adopted is the same across the whole manuscript: dark blue circles: main electronic shock front; electric blue downward triangles: secondary electronic shock front; light blue squares: third electronic shock front; dark red stars: main thermal shock front; light red rightward triangles:  secondary thermal shock front; pink leftward triangles: third thermal shock front.}
    \label{fig:dyn}
    \end{figure*}

	The dynamics of the collision, as well as the different emerging times of the multiple fronts, are well captured by the evolution of the intensities of the fronts at $P=1.0$ W, reported in panel a) of Fig. \ref{fig:dyn}.  Here we clearly see how the different peaks emerge at different times - with the electronic ones appearing faster than the temporal resolution, while the thermal ones emerging on a different timescale later into the dynamics ($t=0.1$ s, $t=0.13$ s and $t=0.2$ s). This also shows how the intensity of the main electronic peak (dark blue) and the secondary electronic peak (blue) drastically drops at $t=0.16$ s and $t=0.22$ s respectively. This happens due to the increasing presence of the thermal main front which gains intensity and collides with the electronic fronts one after the other, slowly prevailing. This transient phase is then followed by a stationary one where both the third electronic front and the main thermal one coexist with equal intensities. In panel b) of  Fig.~\ref{fig:dyn} we report the position of the main peaks throughout the evolution. The behavior during the transient phase is characterized by an initial abrupt beam expansion driven by the combined electronic and thermo-optical defocusing effect. After reaching a maximum, such expansion is arrested, the growth trend is inverted, and the beam undergoes a contraction leading to a stationary size. Saturation effects in the expansion of DSWs have been reported in many systems such as semiconductor-doped glass \cite{Coutaz91}, photo-refractive media \cite{Kivshar03} and Rb vapor \cite{Pavloff21}. To the best of our knowledge, a contraction of a DSWs has never been observed. We ascribe this behavior to the interaction between the local electronic response with the nonlocal thermo-optical one. Indeed, accordingly with the non-linear coefficients reported in the literature~\cite{Sou08} with  $n_2^{(th)}/n_2^{(el)}=10^{7}$  we expect that although the thermo-optic effect responds over longer times, once the thermal lensing is formed, it becomes dominant with respect to the shock front produced by the electronic nonlinearity. The consequences of such thermal lensing supremacy with respect to the electronic nonlinearity are twofold: i) the contraction of the overall shock phenomenon observable during the transient phase; ii) the damping and definite extinction of the oscillations of the electronic shock front close to the thermo-optical shock. Since a theoretical model is still to be developed, further investigation and an ad hoc theoretical model is required to address the emergence of such new phenomenology.\\
	
\begin{figure*}
	\includegraphics[width=\textwidth]{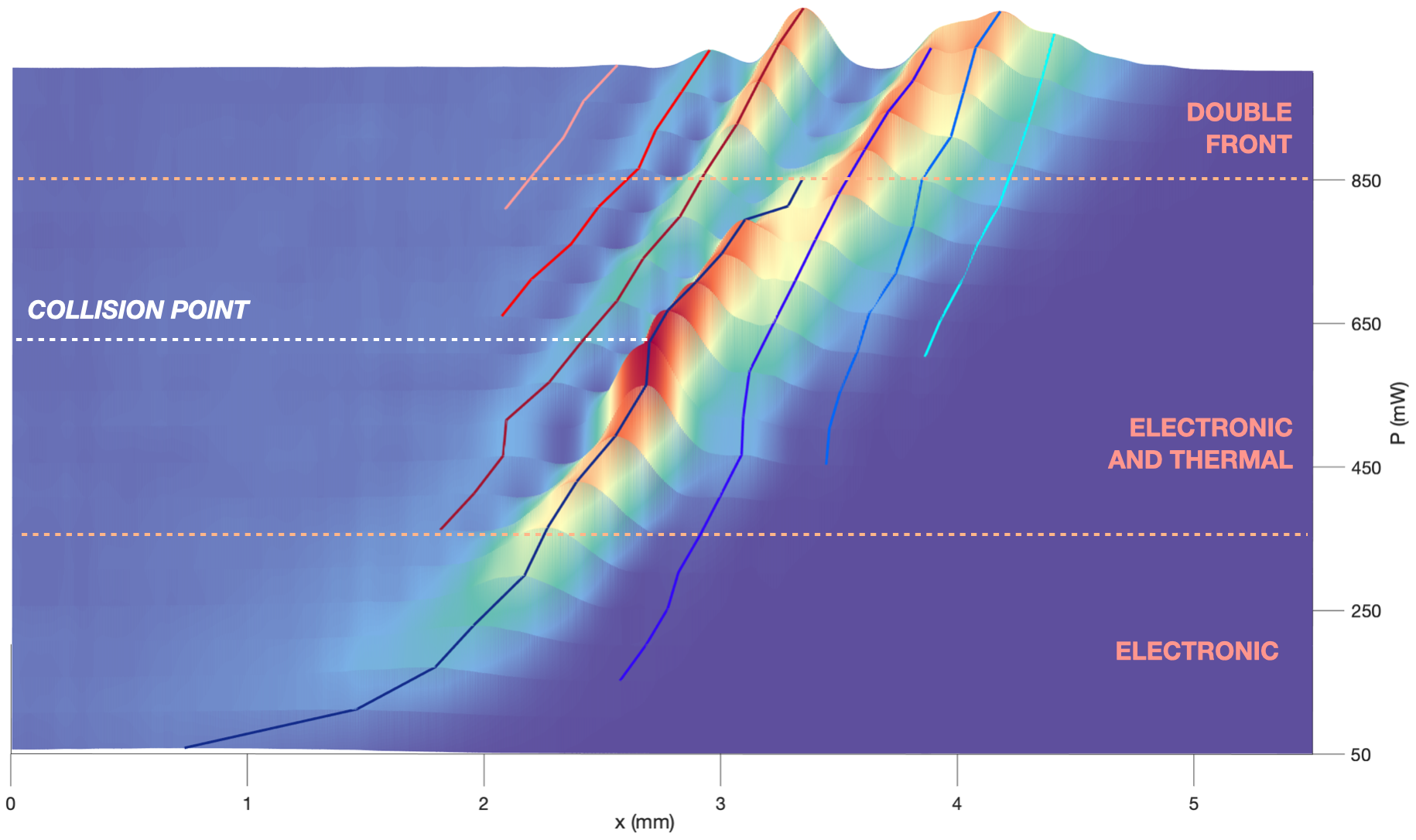}
	\caption[width=\textwidth] {Power-space ($P$-$x$) map obtained by analyzing the intensity profiles collected at varying incident beam power $P$. The continuous lines represent a guide to distinguish the emergence and the evolution of two different shock fronts with the increase of $P$: the continuous blue lines indicate the electronic fronts; the continuous red lines the shock fronts of thermal origin. 
	The orange dashed lines in the map distinguish three different regimes/regions characterized by a different interplay between the two shock fronts. The collision point occurs at $\simeq530$ mW after the main electronic shock front reaches its maximum intensity.}
	\label{fig:Pmaps}
	\end{figure*}

\begin{figure*}[h!]
		\includegraphics[width=\textwidth]{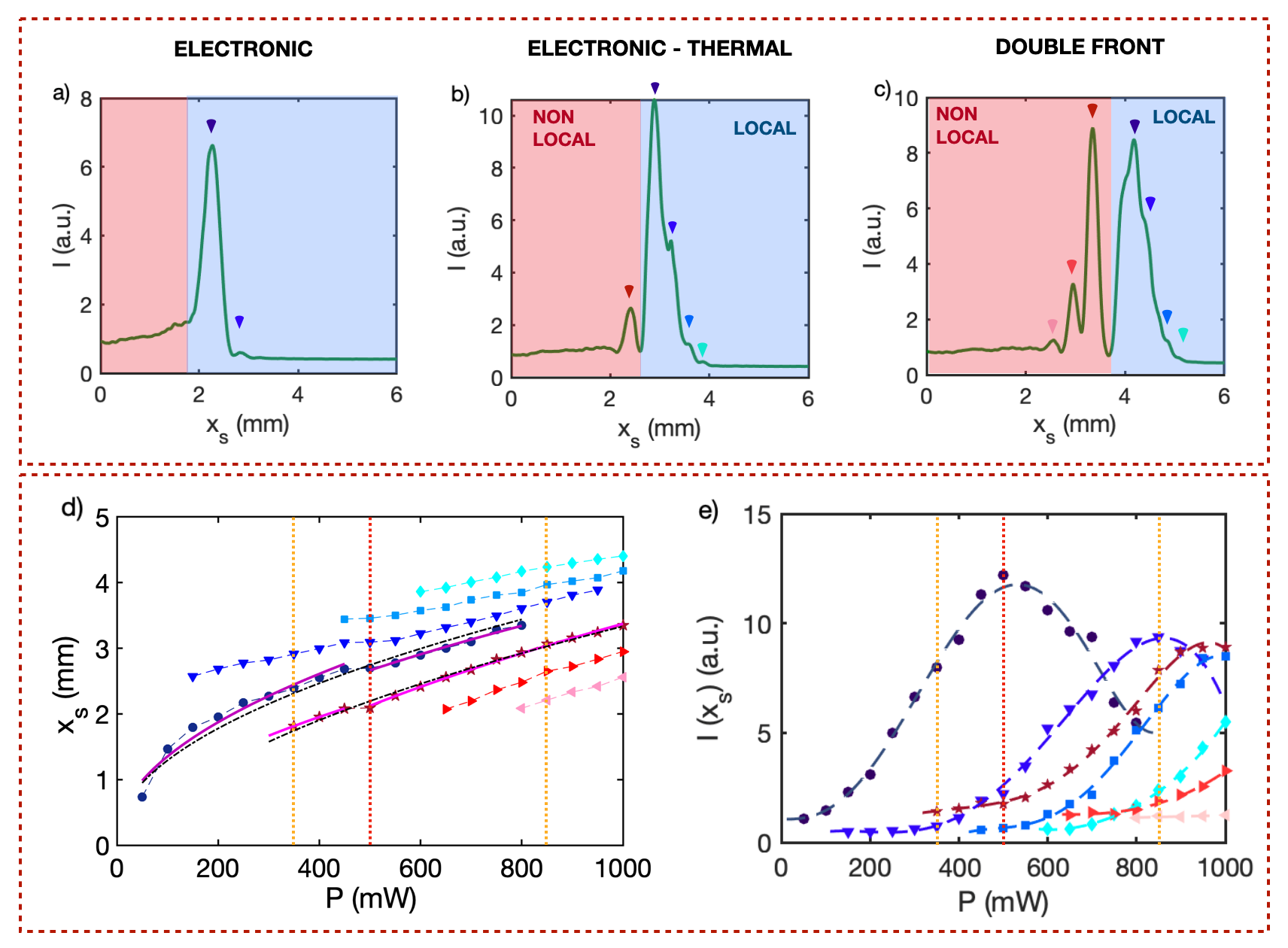}
		\caption {Characterization of the DSWs edges as a function of the incident power, $P$. (a-c) Intensity profiles extracted from the map of Fig.\ref{fig:Pmaps} at $P=300$mW (a),  $=750$mW (b) and $1$W (c). (d) Position $x_s$ of the maxima of the undular bores  versus $P$. The black dot-dahsed lines are the best fit for the two main peaks respectively, considering the whole range of powers according to the function $x_s=a(1+b\cdot P^{1/2})$. The solid purprle and solid fuchsia lines are the best fit considering only the data before and after $P=500$ mW. The vertical red dotted lines highlights the value of $P=500$ mW, while the vertical dotted orange lines indicate the regions defined in the map of Fig.\ref{fig:Pmaps} (electronic, electronic and thermal, double front). (e) Intensity $I(x_s)$ of the maxima of the undular bores versus $P$. The vertical dotted orange lines indicate the regions defined in the map of Fig.\ref{fig:Pmaps} (electronic, electronic and thermal, double front). In panels (d-e) the color scheme is the following: dark blue circles: main electronic shock front; electric blue downward triangles: secondary electronic shock front; light blue squares: third electronic shock front; cyan diamonds: fourth electronic shock front; dark red stars: main thermal shock front; light red rightward triangles:  secondary thermal shock front; pink leftward triangles: third thermal shock front }
		\label{fig:Pcuts}
	\end{figure*}

	{\bf \noindent Shocks evolution by increasing nonlinearity.}
    We then perform the measurements of the spatial profiles as a function of the laser input power $P$ varying between 50 mW and 1.0 W, ensuring that at each power the spatial profiles reaches its stationary state before recording the measurement. Fig. \ref{fig:Pmaps} shows a three dimensional power-space ($P$-$x$) map of the intensity of the half-beam spatial profile averaged over 10 pixel rows, as before.
    We identify three distinct regions in the profile map: a) {\it Electronic (50 mW to 300 mW)}: low powers are characterized by the presence of a main shock front which is soon followed by external oscillations. We thus attribute this behaviour to a local response, as expected.
    {\it b) electronic and thermal (300 mW to 850 mW):} as the power (and hence the nonlinearity) increases, multiple internal and external oscillations appears, with a more involved dynamics unfolding at higher powers. We categorize the internal bores as nonlocal (red shades) and the external ones as local (blue shades). 
    {\it c) double front (from 850 onwards)}: starting from $P=$ 850 mW, as the main thermal front gains strength and draws closer to the main electronic one, the latter starts to drop, giving way to the next electronic front, which, in turn start to decrease as the next one takes its place. During this process the intensity of the main (varying) electronic front and that of the thermal one are comparable, hence we refer to this regime as a double-front.For the sake of clarity, we report in  Fig.~\ref{fig:Pcuts} (a-c) the representative profiles for each region in the $P$-$x$ map of Fig.~\ref{fig:Pcuts}.

    In order to fully capture the interaction between the local and nonlocal DSW, we calculate the position $x_s$, and the intensity $I(x_s)$ of the maxima of each detected peak as a function of the beam intensity $P$. The result of the analysis are reported in Fig.~\ref{fig:Pcuts} (d) and (e). 
    Panel (d) shows the position of the four electronic and three thermal bores generated at increasing power. This allows to verify that the main electronic and thermal fronts (dark blue circles and dark red stars) abide by the expected power law in the hydrodynamic limit of the NLSE. Taking the shock width as a measure of its velocity, it is shown that the front speed increases with intensity ratio according to the relation 
     $x_s=a(1+b\cdot P^{1/2})$ \cite{Bar07,Wan07}. The best fits of the two main fronts are shown as black dot-dashed lines. We note however that, for all fronts, there is a discontinuity occurring at $P=500$ mW (highlighted by a dotted red line on the plot), which visibly leads to two different slopes before and after this value. We thus also evaluate the best fit for the two separated sets. This allows us to relax the constraint on the fit of the electronic front. While exploring the dynamics in temporal domain, we identified the behaviour at $P=500$ mW as the onset of the collision process. The discontinuity in the fronts peak position seems to corroborate this hypothesis, which finds further confirmation when looking at the intensities of the fronts, shown in panel (e) of  Fig.~\ref{fig:Pcuts}:  the intensity of the main electronic peak, initially increasing with the power, reaches its peak at $P=500$ mW and then starts to drop. Even without a full theoretical treatment, the experimental evidence both in the temporal and power domains, solidly points to a collision around $P=500$.  
    Panel (e) also shows how the collision process between the thermal and the main electronic shock front culminates at $P=850$ mW with the extinction of the latter and the growth of the secondary electronic front. From this point onward the intensity of the two phenomena becomes comparable, thus we label this region "double front". Further increasing the power leads to the progressive collision between the main thermal and the highest electronic shock front, which then drops as the next one emerges.

    \section*{Discussion}
    
     In conclusion, the present work reports on the experimental study of a novel optical wave-breaking phenomenon resulting from the interaction of two shock fronts in a dispersive medium displaying a local and a nonlocal defocusing nonlinearity simultaneously. The emerging scenario is a strong interaction between the two DSWs that, at high nonlinearity, leads to the formation of a double front.
     The experimental evidence is solid, but it would benefit from a full theoretical description, which is well beyond the scope of this work. In particular, the model must account for the pulsed structure of light, adding another layer of complexity to the description of the monochromatic paraxial wave equation. This can provide guidance with the analysis not only in the spatial domain, but  in the time-frequency one as well.
     
     On the one hand, our findings provide new insight from a fundamental perspective, allowing for the exploration for the first time of two DSWs together. On the other hand, our results bear strong repercussions for applications: 
     as nonlinear effects in novel materials, including soft matter and biological suspensions, are being explored as new avenues for monitoring and sensing, enhancing their response through AuNR allow for lower-power, less invasive probing. Our findings provide evidence that the resulting behaviour is both intricate and intriguing: tuning the power varies not only the intensity of the nonlinear response, but even its nature. This effect ought to be accounted for in current applications, while, at the same time, it may be harnessed for developing unprecedented ones.

	
	\section*{Methods}
	\noindent
	{\bf Dispersive shock waves.} 
	The spatial propagation of the laser beam in a lossless nonlinear medium is well described by the paraxial wave equation for the slowly varying envelope $A$ of a monochromatic field $E=(2/c\epsilon_0n)^2A\exp(ikZ-i\omega T)$ with intensity $I=|A|^2$:
\begin{equation}
    i\frac{\partial A}{\partial Z}+\frac{1}{2k}\nabla^2_{X,Y} A+k_0\Delta n A=0
\label{eq1}\end{equation}
where $n_0$ is the linear part of the refractive index, $k_0=\frac{\omega}{c}n_0$, and $\Delta n=n_2 I$ is the nonlinear refractive index variation of Kerr type. Eq. (\ref{eq1}) can be rewritten as Eq.\eqref{eq.NLS} in the main text by normalising the field to the peak intensity $I_0$, $\psi=A/\sqrt{I_0}$, and introducing the dimensionless coordinates $\{x,y,z\}=\{X/w_0,Y/w_0,Z/L\}$. The characteristic length scales $L_d=kw_0^2$,  the diffraction length of a beam with waist $w_0$, and $L_{nl}=n_0/(k|n_2|I_0)$ the nonlinear length; finally, $L=\sqrt{L_dL_{nl}}$, thus we can define the parameter $\varepsilon=L_{nl}/L=\sqrt{L_{nl}/L_d}$. 

In the strongly nonlinear regime, i.e. when $\varepsilon\ll1$, Eq. (\ref{eq1}) can be further reduced to a pair of coupled equations describing the behavior of the optical field as that of a fluid undergoing extreme conditions. For this reduction, we consider as the initial condition an axi-symmetric Gaussian beam $\psi_0(r)=\exp{(-r^2)}$ with $r=\sqrt{x^2+y^2}$, and look for a solution in the framework of WKB transformation, i.e. $\psi(r,z)=\sqrt{\rho(r,z)}\exp[i\phi(r,z)/\varepsilon]$, where $\rho(r,z)$ and $\phi(r,z)$ are the intensity and the coherent phase of the optical field, respectively~\cite{WKB}. Operating such a solution in Eq. (\ref{eq.NLS}) and retaining only the leading orders in $\varepsilon$, we obtain Eqs. (\ref{eq.hydrodynamic1}) and (\ref{eq.hydrodynamic2}) in the main text, as the one-dimensional limit.




	\noindent{\bf Samples preparation.} AuNRs synthesis is based on the seed mediated methods and consist in two steps. At first, the seed solution is prepared and then, in the second step, silver nitrate, auric tetrachloric acid, seed solution and ascorbic acid are mixed to grow the rods, as reported in our recent publication \cite{Venditti21}. The purification was performed by centrifugations (13000 rpm, 10 minutes, 2 times) and Uv-Visible spectrum was acquired by using a quartz cuvette and a double-beam Shimadzu UV–2401PC UV–Vis Recording Spectrophotometer, in the wavelength range 800–190 nm. The typical surface plasmon resonance (SPR) bands at 516 and 695 nm (see Supporting) appear and the nanodimensions were confirmed by transmission electron microscopy (TEM) showing dimensions 15-45 nm and aspect ratio 3~\cite{Venditti21}. Always following this recipe, more syntheses were made and the studies were repeated on the different batches.
	In order to perform the experiment, the AuNRs are prepared in an aqueous solution with a concentration of 1.6 mg/mL which is used for all the measurements here presented.

	\noindent {\bf Experimental details.} The characterization of the shock wave spatial profiles is performed using a tunable Ti:Saph laser (Chameleon Ultra II - Coherent) centered at 807 and a repetition rate of 80 MHz. The laser power is adjusted using a Half wave plate (HWP) and a polarization beam splitter (PBS), varying from 50 mW to 1W at the sample.
	In order to have Fourier limited pulses, these are compressed by means of a double-prism compressor. We have checked the compression using a home-built SPIDER device. The SPIDER setup picks off the beam before the sample and the test pulse and its delayed replica are obtained with the front and back reflections of a microscope slide. The part of the beam transmitted through the slide serves as the ancillary beam, which is then stretched using a double grating geometry with two reflective gratings with 1200 g/mm positioned 10 cm apart. The ancillary beam and the two test pulses are upconverted through a 100 $\mu$m BBO Type II crystal and the generated signal is isolated by means of bandpass filters and recorded with a spectrometer (Flame S - Ocean Insight). The SPIDER measurement has reported a factor of 1.28 on the temporal FWHM with respect to the transform limited pulses. This is due to a residual third order phase. See the supplementary materials for more details. The resulting pulses have thus a duration of 159 fs. After the compressor the collimated beam with a diameter of 3 mm is sent on the sample using low-dispersion mirrors. The sample is placed in a 0.8 cm wide couvette (Hellma) with a 2 cm path length. The beam is then collected on a monochrome Retiga CCD camera with 30 fps frame Rate at full resolution placed at 70 cm from the sample. Absorptive ND filters are used to attenuate the beam. Absorptive filter have been chosen rather than reflective ones to avoid imaging artifacts.\\

\bibliography{shock}

\vskip 5 mm

{\bf Acknowledgements} We acknowledge E. Marconi and L. Tortora for providing support in the characterization of the AuNRs solution, and V. Berardi for fruitful discussion. 
This work was supported by the European Commission (FET-OPEN-RIA STORMYTUNE, Grant Agreement No. 899587).\\

{\bf Author contributions} M.B. N.G. and I.G. conceived the project, I.G. and N.G. performed the experiment, S.G. performed the analysis with inputs from N.G. and I.G,  I.V. and C.B. synthesised and characterised the AuNR, N.G. and M.B supervised the project. All authors discussed the results and agreed on their interpretation. The manuscript was initially prepared by I.G. and S.G. with inputs from N.G. and M.B.\\

{\bf Competing financial interests.}
The authors declare no competing financial interests.




	\end{document}